\documentclass[aps, prl, amsmath, showpacs, preprintnumbers, superscriptaddress, twocolumn, amssymb, floatfix]{revtex4-1}
\usepackage{graphicx}
\usepackage{mathptmx, textcomp}
\usepackage[latin1]{inputenc}

\begin{document}
\title{A single ion as a three-body reaction center in an ultracold atomic gas}
\author{Arne H\"arter}
\author{Artjom Kr\"ukow}
\author{Andreas Brunner}
\author{Wolfgang Schnitzler}
\author{Stefan Schmid}
\author{Johannes Hecker Denschlag}
\affiliation{Institut f\"ur Quantenmaterie, Universit\"at
 Ulm,
 89069 Ulm, Germany}

\date{\today}

\begin{abstract}
We report on three-body recombination of a single trapped
$\textrm{Rb}^+$ ion and two neutral Rb atoms in an ultracold atom
cloud. We observe that the corresponding rate coefficient $K_3$ depends on collision energy
and is about a factor
of 1000 larger than for three colliding neutral Rb atoms. In the three-body recombination
process large energies up to several $0.1\:\textrm{eV}$ are released leading
to an ejection of the ion from the atom cloud. It is sympathetically
recooled back into the cloud via elastic binary
collisions with cold atoms. 
Further, we find that the final ionic product of the three-body processes
is again an atomic Rb$^+$ ion suggesting that
the ion merely acts as a catalyzer, possibly in the
formation of deeply bound Rb$_2$ molecules.
\end{abstract}

% Scattering of atoms and molecules, Chemical reactions,    Atom, molecule, and ion cooling methods, Ion cooling

\maketitle Early on in the quest for ultracold quantum gases
three-body recombination played a  crucial role as a limiting
factor for Bose-Einstein condensation. It was first investigated
in spin-polarized hydrogen \cite{Hess1983} and somewhat later for
 alkali atoms \cite{Esry1999, Burt1997}. For samples with large scattering lengths
three-body processes can be resonantly modified through the formation of Efimov states \cite{Fer2010}.
 Recently, three-body recombination was investigated with single atom resolution
  \cite{Spethmann2012}.
Combining ultracold atoms with cold trapped ions is an emerging field where
large scattering cross sections naturally come into play due to the comparatively
long range 1/r$^4$ polarization interaction potential. Two-body collisions
between atoms and ions in the low energy regime have been recently studied \cite{Cote2000a, Vul2008, Zip2010, Smi2010,
Hall2011,Rellegert2011,Ran2011}. In this letter, we report on three-body
collisions involving two ultracold $^{87}$Rb atoms and a $^{87}\textrm{Rb}^+$ ion
at mK temperatures.

The ion in our experiment can be regarded as a reaction center, facilitating molecule formation through its large interaction radius. For the work presented here, it is essential that we work with ions and atoms of the same species. This renders charge transfer reactions irrelevant, which otherwise would strongly constrain our measurements.
Since no accessible optical transition is available for Rb$^+$, it is not amenable to laser-cooling and cannot be imaged. We therefore detect the ion and investigate its dynamics in an indirect way, i.e. through its action on the atom cloud.  In our experiments, we place a single ion into the center of an atomic sample resulting in a continuous loss of atoms due to elastic atom-ion collisions. This behavior is interrupted when a highly energetic three-body process ejects the ion from the atom cloud. By examining the statistics of ion-induced atom loss in hundreds of repetitions of the experiment, we can investigate a number of important details of the three-body process, such as its quadratic density-dependence, the energy that it releases, its rate coefficient $K_\textrm{3}$ and the dependence of $K_\textrm{3}$
on collisional energy. We also obtain information on the reaction products.
As an important byproduct, our measurements also demonstrate sympathetic cooling of an ion from eV energies down to about 
$1\:\textrm{mK}$ using an ultracold buffer gas.

The atom-ion collision experiments are conducted in a hybrid
apparatus (for details see \cite{Smi2012}) where a single
$^{87}$Rb$^+$ ion, trapped in a linear Paul trap, is brought in
contact with an ultracold cloud of spin polarized
$^{87}\textrm{Rb}$ atoms ($F=1, m_F=-1$). The atom
cloud is previously prepared at a separate location from where it
is transported to the Paul trap and loaded into a far off-resonant crossed 
optical dipole trap. The dipole trap is
at first spatially separated from the trapped ion by about $50\:\mu$m. To 
start the atom-ion collision experiments it is then
centered on the ion with $µ\textrm{m}$ precision within a few $100\:\textrm{ms}$.
 At this point the atom cloud consists of
$N_{\textrm{at}}\approx 4.0\times 10^4$ atoms at a temperature of
$T_{\textrm{at}}\approx 1.2\: µ\textrm{K}$ and a peak density
$n_\textrm{at} \approx 1.1\times 10^{12}\:\textrm{cm}^{-3}$. At 
trapping frequencies of $(190,198,55)\,\textrm{Hz}$ this results
in a cigar shaped cloud with radial and axial extensions of 
about $10\:µ\textrm{m}$ and $35\:µ\textrm{m}$, respectively.

The single Rb$^+$ ion is confined in a Paul trap driven at
a frequency of $4.17\:\textrm{MHz}$ resulting in radial and axial trapping
frequencies of $350\:\textrm{kHz}$ and $72\:\textrm{kHz}$, respectively. 
As the trap is about $4\:\textrm{eV}$ deep, the
ion remains trapped on timescales of days and can typically be
used for thousands of experimental cycles. It is initially produced by
photoionization of an atom from a cold Rb cloud in the Paul trap
\cite{tobepub}.
Typical kinetic energies $E_\textrm{ion}$ of the ion after
sympathetic cooling in the atom cloud are about a few mK$\cdot k_\textrm{B}$
where $k_\textrm{B}$ is the Boltzmann factor. This energy scale is mainly
set by two quantities: (1) The excess micromotion (eMM) \cite{Berkeland1998} in the Paul trap 
whose main part we can control by compensating stray electric fields \cite{tobepub}. 
(2) Heating effects induced by the interplay of micromotion and elastic collisions 
\cite{DeVoe2009,Zip2011,Cetina2012}.

\begin{figure}
\includegraphics[width=8.0cm] {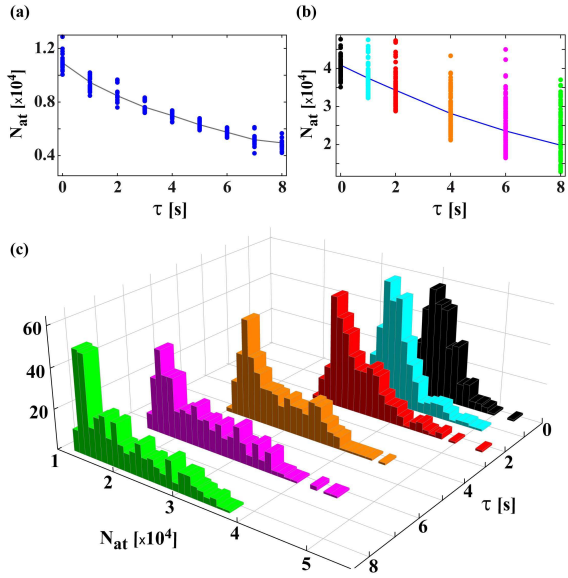}
 %\vspace{-5mm}
\caption{ Decay of the atom cloud under influence of a single trapped ion.
(a) Remaining atom numbers after interaction time $\tau$ for an ion with 
$E_\textrm{ion}\approx 35 \:$mK$\cdot k_\textrm{B}$ \cite{median} and 
$n_\textrm{at}\approx 10^{11}\:\textrm{cm}^{-3}$.  
The solid line indicates the decay of the mean atom number.
(b) Same as (a) but $E_\textrm{ion}\approx 0.5 \:$mK$\cdot k_\textrm{B}$ \cite{median} and 
$n_\textrm{at}\approx 10^{12}\:\textrm{cm}^{-3}$.
(c) Histograms of the data shown in (b). The vertical axis counts the number of
incidents for a particular atom number $N_{\textrm{at}}$ and interaction time $\tau$.
}
 \label{fig1}
\end{figure}

\begin{figure}
\includegraphics[width=8.0cm] {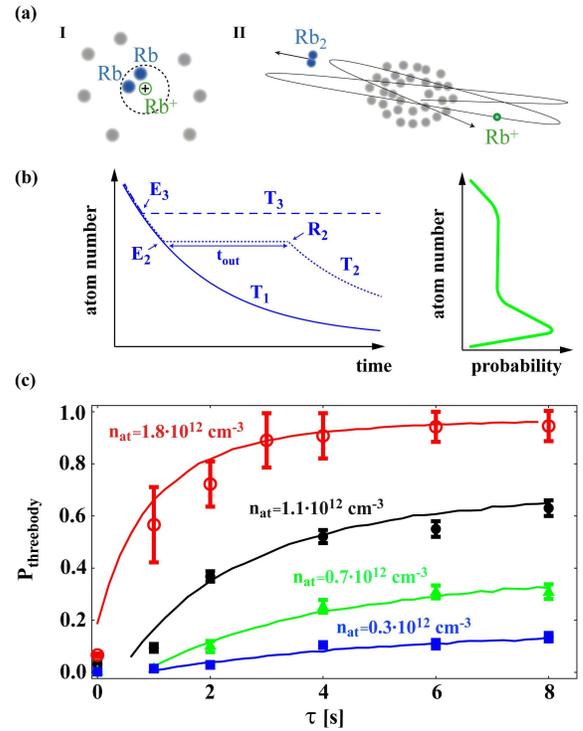}
 %\vspace{-5mm}
\caption{(a) Illustration of an atom-atom-ion collision.
(I) Two atoms simultaneously enter the interaction radius of the ion and a three-body process takes place.
(II) The three-body reaction ejects the ion onto a trajectory much larger than the atom cloud.
 (b) Illustration of our simple model.
\textit{Left:} Various possible time traces for the atom number.
If only binary atom-ion collision occur the atomic sample decays
exponentially (Trace T$_1$). Three-body events (E$_2$, E$_3$) interrupt the atom loss until
the atom is recooled and reenters the sample at point R$_2$ (Traces T$_2$ and T$_3$).
\textit{Right:} Sketch of the resulting atom number distribution when averaging over many experimental runs.
(c) Plot of the probability $P_\textrm{threebody}$ for initial atomic densities 
$(1.8, 1.1, 0.7, 0.3)\times 10^{12}\textrm{cm}^{-3}$ and atom numbers $(6.5, 4.0, 2.8, 1.6)\times 10^{4}$,
 respectively. The solid lines are results of the numerical simulation.}
 \label{fig2}
\end{figure}

As described in  \cite{Smi2010}, an ion immersed in an ultracold atom cloud leads to atom loss by expelling atoms from the shallow optical trap ($\approx 10\:\mu \textrm{K} \cdot k_\textrm{B}$ trap depth) via elastic collisions. The radio frequency (rf) driven micromotion
 is a constant source of energy which drives these loss-afflicting collisions. Figure$\,$\ref{fig1}a shows such a decay of an atom cloud at relatively low densities ($\approx$10$^{11} \:$cm$^{-3}$) and relatively high ion energies ($\approx 35\:\textrm{mK}\cdot k_\textrm{B}$ \cite{median}). Plotted is the number of remaining atoms after an atom-ion interaction time $\tau$. Each data point corresponds to a single measurement.  Overall, the plot shows a relatively smooth decay of the atom cloud with a relative scatter of the atom number of less than $10\%$. This changes drastically when we carry out the experiments at low ion energies ($\approx$0.5 mK$\cdot k_\textrm{B}$ \cite{median}) and larger densities ($\approx$10$^{12} \:$cm$^{-3}$) (Fig.$\,$\ref{fig1}b).
Here, the scatter dramatically increases with $\tau$ and is on the order of the number of lost atoms.
In Fig.$\,$\ref{fig1}c histograms of the data in Fig.$\,$\ref{fig1}b are shown. With increasing time $\tau$
the initial normal distribution becomes bimodal as a striking tail towards large atom numbers emerges. At the tips of the tails we find cases where even after interaction times of several seconds barely any atoms have been lost, a 
signature of missing atom-ion interaction. Apparently, sporadically the ion is ejected from the atom cloud
and promoted onto a large orbit for a period of time during which atom-ion collisions are negligible (Fig.$\,$\ref{fig2}a).
In principle, this is reminiscent of the energy distributions with high energy tails that have recently been predicted 
for trapped ions immersed in a buffer gas \cite{DeVoe2009,Zip2011}.
However, it turns out that such an explanation is inconsistent with our observations \cite{powerlaw}. Rather, we find that it is a three-body recombination process involving the ion and two neutrals that ejects the ion from the cloud. If, for example, a ground state molecule is formed, binding energies on the order of $0.5\:\textrm{eV}$ can be released. Due to the large trap depth the ion is not lost in such an event, but it is recooled back into the cloud through binary collisions after some time.

Figure \ref{fig2}b illustrates in a simple picture how the decay of the atom number over time can follow different paths. The solid trace T$_1$ shows the case when only binary atom-ion collisions occur. Such traces result in the narrow peak of the atom number distribution sketched on the right of Fig.$\,$\ref{fig2}b. Traces T$_2$ and T$_3$ exhibit three-body collisions at points E$_2$ and E$_3$. At point R$_2$ the ion reenters the atom cloud after recooling. Rare three-body events and long interruption times $t_\textrm{out}$ result in a long tail of the distribution. We can reproduce the histograms in Fig.$\,$\ref{fig1} with a simple Monte Carlo type simulation. We assume an initial normal distribution of the atom number which then decays exponentially with the binary atom-ion collision rate $K_2 n_{\textrm{at}}$. Here, $K_2 \approx 5 \times 10^{-9}\:\textrm{cm}^3/ \textrm{s}$ is a rate constant given by the product of the elastic cross section and the ion velocity. A three-body event, occurring at a rate $K_3 n_{\textrm{at}}^2$, interrupts this decay for a period $t_\textrm{out}$. As the ion can only be recooled by the atomic sample, we assume the rate for reentry of the ion into the atom cloud to be proportional to the number of atoms $1/\left\langle t_\textrm{out}\right\rangle= N_\textrm{at}/c_\textrm{out}$ with $c_\textrm{out}$ being a
constant depending on the trap parameters. In order to fix the model parameters and rate constants it is convenient to carry out a data analysis as follows.
We sum up the number of events in the tail of the distributions shown in Fig.$\,$\ref{fig1} for each $\tau$ and divide by the total number of measurements  to obtain the probability $P_\textrm{threebody}$ that at least one three-body process takes place within time $\tau$.
These data are plotted as the filled black circles in Fig.$\,$\ref{fig2}c (data set with $n_\textrm{at}\approx 1.1\times 10^{12}\:\textrm{cm}^{-3}$).
 The corresponding errorbars mainly reflect the difficulty to unambiguously assign
 data points to the peak or the tail of the distribution where both parts of the bimodal distribution
strongly overlap.
Besides the data of Fig.$\,$\ref{fig1}c, Fig.$\,$\ref{fig2}c  also contains data
at three additional atomic densities. All four data sets have in common that
the number of three-body events first rapidly increases and subsequently levels off.
The leveling off is due to the fact that the probability for a three-body reaction is strongly density-dependent.
Closer inspection of the data sets reveals a peculiar feature. In the beginning of the interaction ($\tau\lesssim 1\,\textrm{s}$) only very few three-body events are detected for the lower density samples. We explain this delay by an initial phase of sympathetic cooling of the Rb$^+$ ion which experiences significant heating during the preparation time of the atom cloud (\texttildelow $30\,\textrm{s}$). In fact, we observe ion heating specifically during evaporative cooling where the rf is ramped from $70$ to $3\:$MHz. From
numerical calculations we estimate that recooling times of about $1\,$s in atom clouds with 
$n_\textrm{at}\approx 10^{12}\:\textrm{cm}^{-3}$ correspond roughly to ion kinetic energies of a few $100\:$K$\cdot k_\textrm{B}$. According to the calculations the ion will typically undergo several thousand binary collisions with cold atoms until it is sympathetically recooled to mK$\cdot k_\textrm{B}$ energies.
We are able to describe all four data sets in Fig.$\,$\ref{fig2}c consistently with our simple Monte Carlo model (continuous lines). From independent fits to each data set we obtain nearly identical rate coefficients in the range  $K_3=3.1-3.5\times 10^{-25}\:\textrm{cm}^{6}/\textrm{s}$. For each fit the initial cool-down time is accounted for by adjusting the starting time of the simulation.
 We note that the value for our atom-atom-ion $K_3$ rate coefficient is more
than three orders of magnitude larger than the three-body coefficient for three colliding neutral $^{87}$Rb atoms \cite{Esry1999}.

In order to challenge our analysis we have attempted to model the events that send the
ion into orbit as two-body processes. The corresponding linear
density dependence of the event rate yields much less consistent
fit results. The two-body rate coefficients would differ by more than a factor of 3 when comparing the analysis of the data set for the lowest and the highest density.
As a cautionary note, we point out that three-body recombination processes to weakly-bound molecular states with binding energies $\lesssim 10\:\textrm{meV}$ are not detected in our experiments as the ion will not leave the atom cloud. Thus, the true three-body coefficient may even be significantly larger.
From our analysis we also  extract an approximate value for the interruption time coefficient
$c_\textrm{out}\approx 1.7\times 10^5 \,\textrm{s}$. For the typical atom numbers used here this results in several seconds of negligible atom-ion interaction following each ejection of the ion.

\begin{figure}
\includegraphics[width=8.0cm] {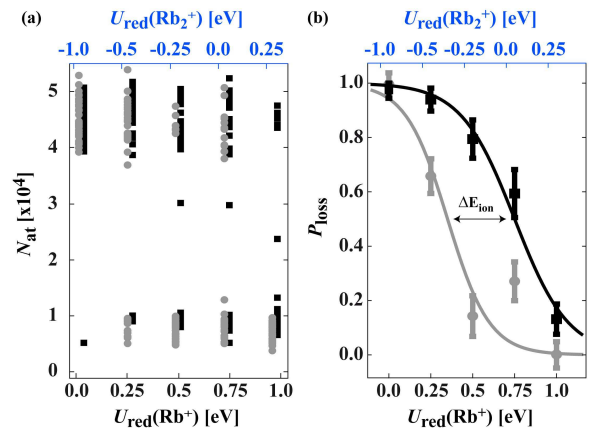}
 %\vspace{-5mm}
\caption{Measuring the ion energy after a three-body process (black squares) and after purely binary collisions (grey dots). The trap depth is reduced to $U_\textrm{red}$. 
(a) If the ion "survives" this procedure in the trap it 
will subsequently induce atom loss in a freshly prepared cloud of atoms. Plotted is the remaining 
atom number. Data points with $N_\textrm{at}\lesssim 10^4$ signal the presence of an ion, 
while data points with $N_\textrm{at} \approx 4.5\times10^4$ correspond to a lost ion.
For better visibility, we have slightly offset in energy the black squares from the grey
dots.
(b) Ion loss probability $P_{\,\textrm{loss}}$ calculated from the data in (a).
The continuous lines are guides to the eye.
}
\label{fig3}
\end{figure}

In a further experiment, we quantify the kinetic energy $\Delta E_\textrm{ion}$ gained by the ion
in a three-body event. The idea is to lower the depth of the ion trap such that
an ion with an energy of a few $0.1\:$eV escapes while a cold ion remains trapped. The trap depth is reduced to one of 5 values $U_\textrm{red}$ by lowering one of the endcap voltages of the Paul trap within $300\,$ms. The voltage is held at this value for $200\,$ms and ramped back up within $200\,$ms. The trap depths $U_\textrm{red}$ are indicated in Fig.\:3. They have been computed using methods detailed in \cite{bem} for both Rb$^+$ (bottom abscissa scale) and Rb$_2^+$ (top). A negative trap depth value corresponds to non-trapping. After the trap depth reduction procedure is completed, the presence of the ion in the trap is probed via
the loss it induces in a freshly prepared atom cloud containing about $5\times 10^4$ atoms. 
For this probing procedure we deliberately apply an offset electric field of about $6\,\textrm{V/m}$
to increase the eMM energy. In this way, we make three-body reactions unlikely and induce a rapid loss of atoms through binary atom-ion collisions.
Figure $\ref{fig3}$a shows the remaining atom number after
$6\,\textrm{s}$ of interaction time. An atom number $\lesssim 1\times 10^4$ indicates the presence of an ion while a number around $4.5\times 10^4$ shows its absence. The clear splitting of the two groups of data allows for ion detection with an efficiency close to unity. In addition, this ion detection is energy resolved. Figure $\,$\ref{fig3}a contains two different plot symbols, distinguishing two classes of ions that have undergone a different prehistory before their energy resolved detection. Black plot symbols correspond to ions that have been promoted to a high energy orbit due to a three-body recombination. The recombination occurred during 4s of atom-ion interaction and is detected through strongly reduced atom loss. Grey plot symbols correspond to ions where no suppression of atom loss was detected. These ions should in general have low kinetic energy. We now analyze the data points of Fig.$\,$\ref{fig3}a by calculating the probability for ion loss $P_{\,\textrm{loss}}$ (see Fig.$\,$\ref{fig3}b).
As expected, ions that were involved in a three-body recombination process can in general escape from deeper traps than cold ions. From the energy offset between the black and grey data we estimate the gained energy $\Delta E_\textrm{ion} \approx 0.4\:\textrm{eV}$.
We note that for trap depths $U_\textrm{red} \lesssim 0.25\:$eV the probability of loss is high in general. This suggests that the stability of our trap is compromised at such shallow trapping potentials, limiting the accuracy with which we can determine the energy released in the three-body process. Still, we find a clear splitting between the black and grey data sets. Thus, a resolution of the measurement on the order of $0.1\:\textrm{eV}$ seems plausible.

Mainly two recombination processes come into consideration. In a reaction of the type Rb + Rb + Rb$^+ \rightarrow$ Rb$_2$ + Rb$^+$ the formation of a neutral molecule is catalyzed by the ion which carries away $2/3$ of the energy released.
If deeply bound Rb$_2$ molecules are produced binding energies of up
to \texttildelow$\:0.5\:\textrm{eV}$ are released, in agreement with the measurement.
A second possible recombination process, Rb + Rb + Rb$^+ \rightarrow$ Rb$_2^+$ + Rb,
produces a molecular ion and a neutral atom. However, as indicated in figure 3, the molecular ion, due to its higher mass, 
experiences a significantly shallower trap than the atomic ion and would immediately be lost for our parameter range.
 We thus infer that the ion at hand is Rb$^+$. However, we cannot completely exclude
the formation of an intermediate molecular ionic state which may subsequently dissociate.

\begin{figure}
\includegraphics[width=8.0cm] {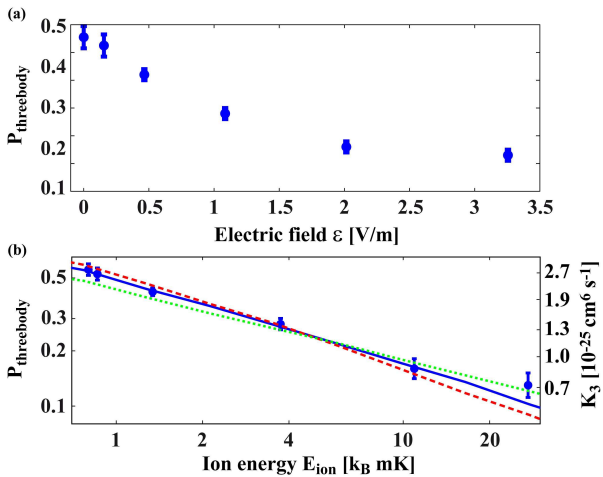}
 %\vspace{-5mm}
\caption{(a): $P_\textrm{threebody}$ as a function of the external electric field.
(b): Double-logarithmic plot of
$P_\textrm{threebody}$ as a function of the ion energy ${E}_\textrm{ion}$ \cite{median}.
A scale for the three-body coefficients $K_3$ as derived from the simulation is also given (see text for details).
}
 \label{fig4}
\end{figure}

In a third type of measurement we study the dependence of the three-body coefficient on the ion kinetic energy which we can tune by controlling the ion micromotion. For this we apply a static electric field $\epsilon$ perpendicular to the axis of the Paul trap and let the ion interact for $\tau=8\,\textrm{s}$ with an atom cloud with $n_\textrm{at} \approx 1.0\times 10^{12}\:\textrm{cm}^{-3}$.
We find $P_\textrm{threebody}$ to increase roughly by a factor of 5 when
reducing $\epsilon$ from $3.25\:\textrm{V/m}$ to $0\:\textrm{V/m}$
(Fig.$\,$\ref{fig4}a).

In order to express the electric field values in terms of kinetic energy
 we make use of the relation $E_\textrm{eMM} = c_\textrm{trap} \cdot \epsilon^2 + E_\textrm{res}$ 
 with $c_\textrm{trap}$ a constant that depends on the trap configuration and the ion mass \cite{Berkeland1998}.
 $E_\textrm{res}$ stands for residual uncompensated micromotion energy, e.g. due to rf phase delay between the trap electrodes. The ion energy can be expressed as $E_\textrm{ion} = c_\textrm{dyn} \cdot E_\textrm{eMM}$ \cite{median}. $c_\textrm{dyn}$ is a constant which depends on the atom-ion mass ratio and the spatial extension of the atom cloud and for our experiments can be estimated to be about 2 \cite{Zip2011}. Since three-body coefficients often follow simple scaling laws, we attempt to describe our data with a power-law dependence of the form $K_3\propto E_\textrm{ion}^\alpha$ within our simulation. We have taken care to also account for the energy dependence of the two-body interactions. Good agreement with the data is achieved for $\alpha=-0.43$, $E_\textrm{res}=370\,µ\textrm{K}$$\cdot k_\textrm{B}$ and a maximal value for $K_3$ of $2.75\times 10^{-25}\:\textrm{cm}^6/\textrm{s}$ (solid trace in Fig.$\,$\ref{fig4}b). For comparison, curves for
exponents $\alpha=-0.5$ and $\alpha=-0.33$  (dashed and dotted traces, respectively)
are shown as well. 
A residual energy $E_\textrm{res}=370\,µ\textrm{K}$$\cdot k_\textrm{B}$ is a reasonable value for our current setup and in agreement with other measurements of ours \cite{tobepub}.

In conclusion, we have studied three-body recombination involving a single trapped ion and two of its parent atoms at collision energies 
approaching the sub-mK regime. With a relatively simple model we can understand the two- and three-body collision dynamics and extract corresponding rate coefficients. We observe an increase of the three-body rate coefficient with decreasing collision energy, a behavior 
that can be expected to become crucial for future experiments targeting even lower temperatures.
 After a three-body event, ion energies on the order of $0.4\:\textrm{eV}$ were measured, indicating that deeply bound molecules have been created.
Since we have not observed Rb$_2^+$ ions, the formation of Rb$_2$ seems probable. The ion would then act as an atomic size catalyzer at mK temperature.
 It is an interesting question whether such catalyzing action could also be observable close to more massive objects such as carbon nanotubes \cite{Gierling2011}.
 Finally, as a byproduct of our investigations, we also observe sympathetic cooling of ions from energies in the 0.1$\:$eV range back to mK temperatures.

The authors would like to thank Kilian Singer, Piet Schmidt, David Hume, Olivier Dulieu and Brett Esry for helpful discussions and information. This
work was supported by the German Research Foundation DFG within the SFB/TRR21.

\bibliographystyle{apsprl}

\begin{thebibliography}{22}
\bibitem [{pow()}]{powerlaw}%
  \BibitemOpen
  \href@noop {} {}\bibinfo {note} {Performing numerical simulations we find
  that the ion is very unlikely to leave the atom cloud even at the high energy
  power-law tail of the energy distribution. More strikingly, we experimentally
  observe the signature of ion ejection only upon minimization of the ion
  micromotion, in stark contrast to the expectations of the quoted heating
  mechanism.}\BibitemShut {Stop}%
\end{thebibliography}

\end{document}